\documentclass{ws-procs9x6}

\begin{document}

\title{Direct Determinations of the
Redshift Behavior of the 
Pressure, Energy
Density, and Equation of State of
the Dark Energy and the Acceleration of the Universe}

\author{Ruth A. Daly}

\address{Department of Physics\\
Berks-Lehigh Valley College\\
Penn State University\\
Reading, PA 19601-6009\\
E-mail: rdaly@psu.edu}

\author{S. G. Djorgovski}

\address{Division of Physics, Mathematics, and Astronomy \\
California Institute of Technology \\
MS 105-24, Pasadena, CA 91125\\
E-mail: george@astro.caltech.edu}

\maketitle

\abstracts{One of the goals of current cosmological
studies is the determination of the expansion and
acceleration rates of the universe as functions of redshift,
and the determination of the
properties of the dark energy that can explain these
observations.  Here the expansion and acceleration rates are
determined directly from the data, without the need for
the specification of a theory of gravity, and without
adopting an a priori parameterization of the form or
redshift evolution of the dark energy.
We use the latest set of distances to SN standard candles
from Riess et al. (2004), supplemented by data on radio
galaxy standard ruler sizes, as described by Daly \& Djorgovski
(2003, 2004).
We find that the universe transitions from acceleration to deceleration
at a redshift of $z_T \approx 0.4$, with the present value of
$q_0 = -0.35 \pm 0.15$.  The standard ``concordance
model'' with $\Omega_0 = 0.3$ and $\Lambda = 0.7$
provides a reasonably good fit to
the dimensionless expansion rate as a function of redshift,
though it fits the dimensionless acceleration rate as a function
of redshift less well.  The expansion and
acceleration rates are then combined with a theory of
gravity to determine the
pressure, energy density, and equation of state of the
dark energy as functions of redshift.
Adopting General Relativity as the
correct theory of gravity, the redshift trends for the
pressure, energy density, and
equation of state of the dark energy
out to $z \sim 1$ are determined, and are found to be 
generally consistent with the
concordance model; they have zero redshift values of
$p_0 = -0.6 \pm 0.15$,
$f_0 = 0.62 \pm 0.05$, and
$w_0 = -0.9 \pm 0.1$.
}

\section{Introduction}

One way to determine the expansion and acceleration rates of the
universe as functions of redshift is through studies of the coordinate
distance to sources at different redshift.  This can be accomplished with 
a variety of techniques including the use of supernovae and radio galaxies
(e.g. Riess et al. 2004; Perlmutter et al. 1999; Riess et al. 1998;
Daly 1994; Guerra \& Daly 1998;
Guerra, Daly, \& Wan 2000).  The techniques fall into
two broad categories: the integral and the differential approaches.

The former, traditional approach involves the integration
of a theoretically predicted expansion rate
over redshift to obtain predicted coordinate distances to different
redshifts; the difference between these predicted coordinate distances
and the observed coordinate distances is then minimized to obtain
the best fit model parameters.  This approach usually
requires the specification
of a theory of gravity (generally taken to be General Relativity; GR) and
a parameterization of the redshift evolution of the dark energy.
Maor, Brustein, \& Steinhardt (2001) and Barger \& Marfatia (2001)
discuss how difficult it is to extract the redshift behavior of the dark
energy using this method.  Some techniques have
been developed to extract the redshift behavior of the dark
energy using the integral method (e.g. Starobinsky 1998; Huterer
\& Turner 1999, 2001; Saini et al. 2000; Chiba \& Nakamura 2000;
Maor, Brustein, \& Steinhardt 2001; Golaith et al. 2001;
Wang \& Garnavich 2001;
Astier 2001; Gerke \& Efstathiou 2002; Weller \& Albrecht 2002;
Padmanabhan \& Choudhury 2002; Tegmark 2002;
Huterer \& Starkman 2003;  Sahni et al. 2003; Alam et al. 2003;
Wang \& Freese 2004; Wang et al. 2004; Wang \& Tegmark; 
Nessier \& Perivolaropoulos 2004;
Gong 2004; Zhu, Fujimoto, \& He 2004; Elgaroy \& Multamaki 2004;
Huterer \& Cooray 2004; Alam, Sahni, \& Starobinsky 2004).

\section{The Methodology}

The differential approach has been investigated by
Daly \& Djorgovski (2003, 2004), and we offer a brief summary
here.  It is well known (e.g. Weinberg 1972; Peebles 1993;
Peebles \& Ratra 2003)
that the dimensionless expansion rate $E(z)$
can be written as the derivative of the dimensionless coordinate
distances $y(z)$; the expression is particularly simple when
the space curvature term is equal to zero.  In this case,
\begin{equation}
\left( { \dot{a} \over a} \right)~H_0^{-1} \equiv
E(z) = (dy/dz)^{-1}~,
\label{eofz}
\end{equation}
where $a$ is the cosmic scale factor, and
Hubble's constant is
$H_0 = (\dot{a}/a)$
evaluated at zero redshift.
This representation follows directly
from the Friedman-Robertson-Walker line element, and does not
require the use of a theory of gravity.  Similarly,
in a spatially flat universe 
(as convincingly demonstrated by CMBR measurements,
Spergel et al. 2003),
it is shown
in Daly \& Djorgovski (2003) that the dimensionless deceleration parameter
\begin{equation}
- \left({\ddot{a} a \over \dot{a}^2}\right)
\equiv q(z) = - [1+(1+z)(dy/dz)^{-1}~d^2y/dz^2]
\label{qofz}
\end{equation}
also follows
directly from the FRW line element, and is independent of
any assumptions regarding the dark energy or a
theory of gravity.  Thus, measurements of the dimensionless
coordinate distance to sources at different redshifts can be used
to determine $dy/dz$ and $d^2y/dz^2$, which can then be used to
determine $E(z)$ and $q(z)$, and these direct determinations
are completely model-independent, as discussed by Daly \&
Djorgovski (2003).

In addition, if a theory of gravity is specified, the measurements of
$dy/dz$ and $d^2y/dz^2$ can be used to determine the pressure, energy
density, and equation of state of the dark energy as functions of
redshift (Daly \& Djorgovski 2004); we assume the standard GR for
this study.
{\it These determinations are completely independent of
any assumptions regarding the form or properties of the dark
energy or its redshift evolution.}
Thus, we can use the data to determine these functions directly,
which provides an approach that is complementary to the standard
one of assuming a physical model, and then fitting the
parameters of the chosen function.

In a spatially flat, homogeneous, isotropic universe with non-relativistic
matter and dark energy
Einstein's equations are
$({ \ddot{a} / a} ) = -{(4 \pi G / 3})~
(\rho_m + \rho_{DE} + 3 P_{DE})$
and
$({ \dot{a} / a} )^2 = ({8 \pi G / 3})~ (\rho_m + \rho_{DE})~,$
where $\rho_m$ is the
mean mass-energy density of non-relativistic matter,
$\rho_{DE}$ is the mean mass-energy
density of the dark energy, and $P_{DE}$ is the pressure of the dark energy.
Combining these equations, we find
$(\ddot{a}/a)=-0.5[(\dot{a}/a)^2~+(8 \pi G)~P_{DE}]$.

Defining the critical density at the present
epoch in the usual way,
$\rho_{oc} = 3H^2_0/(8 \pi G)$, it is easy to show that
$p(z) \equiv ({P_{DE}(z) / \rho_{oc}}) =
({E^2(z) / 3})~[2q(z)-1]~.$
Combining this expression with eqs. (1) and (2) we
obtain the pressure of the dark energy as a function of redshift
in terms of first and second derivatives of the dimensionless
coordinate distance $y$ (Daly \& Djorgovski 2004)
\begin{equation}
p(z) = -(dy/dz)^{-2}[1+(2/3)~(1+z)~(dy/dz)^{-1}~(d^2y/dz^2)]~.
\label{pofz}
\end{equation}
Thus, the pressure of the dark energy can be determined
directly from measurements of the coordinate distance.
In addition, this provides a
direct measure of the cosmological constant for
Friedmann-Lemaitre models since in these
models $p = -\Omega_{\Lambda}$.  If more than one new component is
present, this pressure is the sum of the pressures of the new
components.

Similarly, the energy density of the dark energy can be obtained
directly from the data
\begin{equation}
f(z) \equiv \left( {\rho_{DE}(z) \over \rho_{oc}} \right)
= (dy/dz)^{-2} - \Omega_{0}(1+z)^3~,
\label{fofz}
\end{equation}
where $\Omega_{0} = \rho_{om}/\rho_{oc}$ is the fractional contribution
of non-relativistic matter to the total critical density at zero redshift,
and it is assumed that this non-relativistic matter evolves as $(1+z)^3$.
If more than one new component is present, then $f$ includes the
sum of the mean mass-energy densities of the new components.

The equation of state $w(z)$ is defined to be the ratio of the pressure of
the dark energy to it's energy-density $w(z) \equiv P_{DE}(z)/\rho_{DE}(z)$.
As shown by Daly \& Djorgovski (2004), the equation of state is
\begin{equation}
w(z) = -{[1 + (2/3)
~(1+z)~(dy/dz)^{-1}~(d^2y/dz^2)] \over [1-(dy/dz)^2~\Omega_{0}~(1+z)^3]}~.
\label{wofz}
\end{equation}
Here, $w$ is the equation of state of the dark energy; if more than one
new component contributes to the dark energy, $w$ is the ratio of the sum
of the total pressures of the new components to their total mean mass-energy
densities.

\begin{figure}[ht]
\centerline{\epsfxsize=3.0in\epsfbox{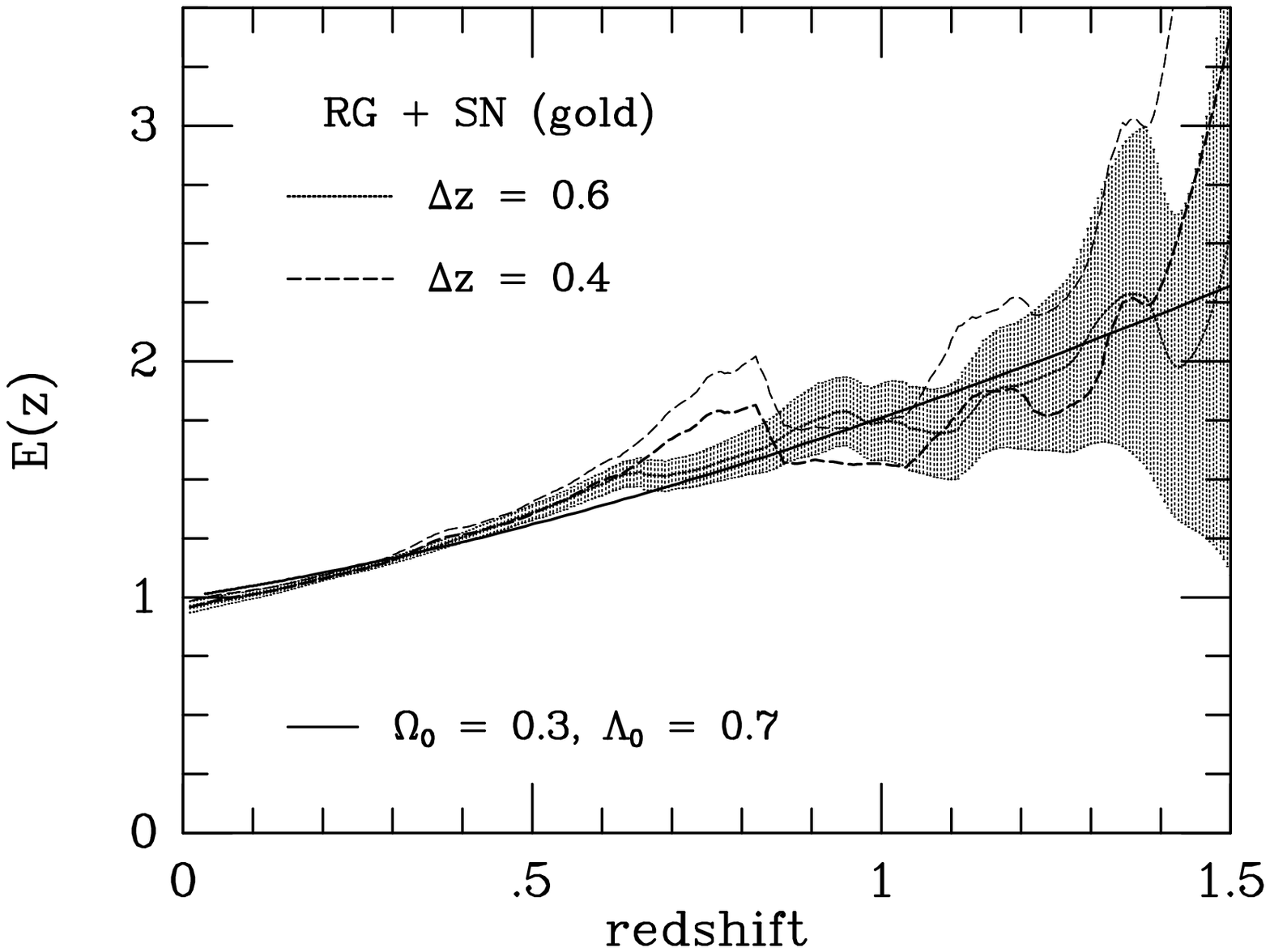}}
\caption{The derived values of the dimensionless expansion rate
$E(z) \equiv (\dot{a}/a)H_0^{-1}=(dy/dz)^{-1}$
obtained with window functions of width $\Delta z = 0.4$
and their 1 $\sigma$ error bars
(dashed lines) and 0.6 (dotted line and hatched error range).
At zero redshift, the value of $E$ is $E_0 = 0.97 \pm 0.03$.
The value of $E(z)$ predicted in a spatially flat universe with
a cosmological constant $\Lambda = 0.7$ and mean mass density
$\Omega_0 =0.3$ is also shown.} 
\bigskip

\centerline{\epsfxsize=3.0in\epsfbox{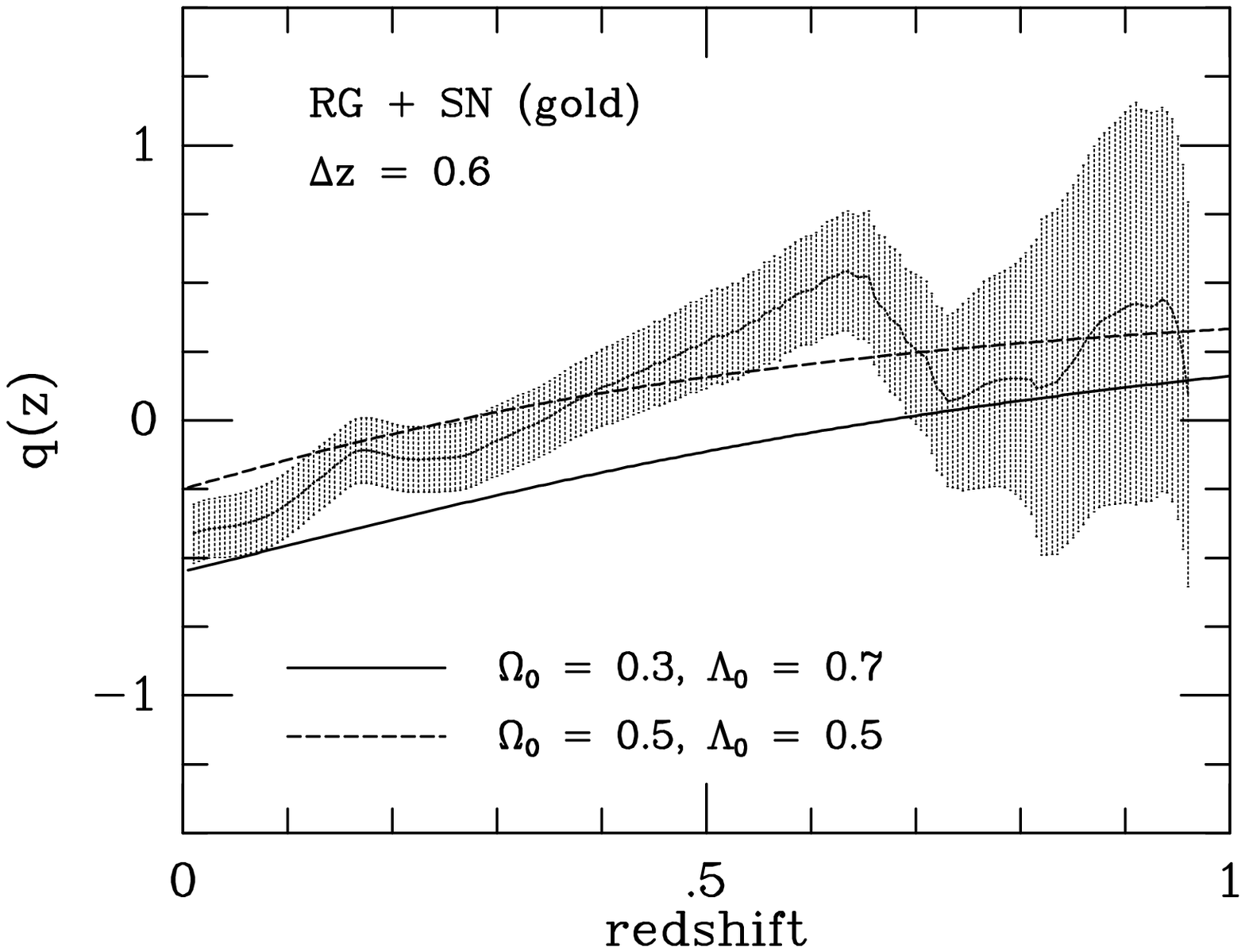}}
\caption{The derived values of deceleration parameter $q(z)$
and their 1 $\sigma$ error bars
obtained with window function of width $\Delta z = 0.6$
applied to the radio galaxy and gold supernovae
samples.  The universe transitions from acceleration
to deceleration at a redshift $z_T \approx 0.4$, with an
uncertainty difficult to quantify due to large fluctuations
at $z > 0.5$, caused by the sparseness of the data at
higher redshifts.
The present value
is $q_0 = -0.35 \pm 0.15$.
Solid and dashed lines show the expected dependence in the standard
Friedmann-Lemaitre models with zero curvature, for two pairs of
values of $\Omega_0$ and $\Lambda_0$.
}
\end{figure}

\begin{figure}[ht]
\centerline{\epsfxsize=3.0in\epsfbox{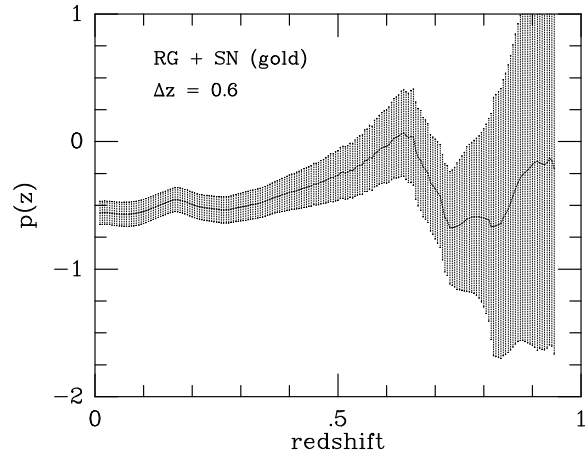}}
\caption{The derived values of dark energy pressure $p(z)$,
obtained with window function of width $\Delta z = 0.6$.
This derivation of $p(z)$ requires a choice of theory of gravity,
and GR has been adopted here.
The present value is $p_0 = -0.6 \pm 0.15$.  Note that in the
standard Friedmann-Lemaitre models, $p_0 = -\Lambda_0$, and
thus we have a direct measurement of the value of the cosmological
constant, which is also fully consistent with other modern
measurements.
}
\end{figure}

\begin{figure}[ht]
\centerline{\epsfxsize=3.0in\epsfbox{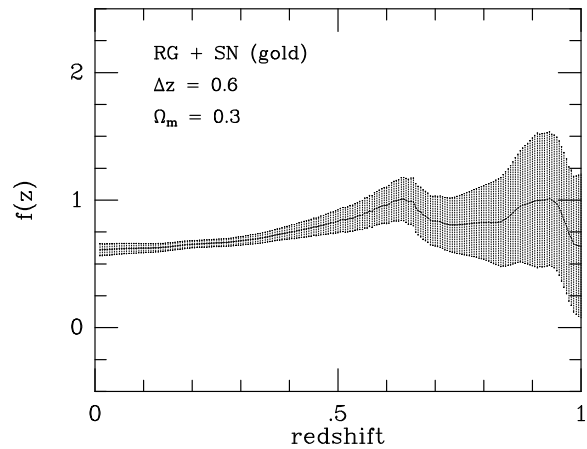}}
\caption{The derived values of the dark energy density fraction
$f(z)$, obtained with window function of width $\Delta z = 0.6$.
This derivation of $f(z)$ requires of theory of gravity
and the value of $\Omega_{0}$ for the nonrelativistic matter;
GR has been adopted here, and
$\Omega_0 = 0.3$ is assumed. The present value is $0.62 \pm 0.05$,
and the trend is consistent with $f(z) = const.$ out to $z \approx 1$.
}
\end{figure}

\begin{figure}[ht]
\centerline{\epsfxsize=3.0in\epsfbox{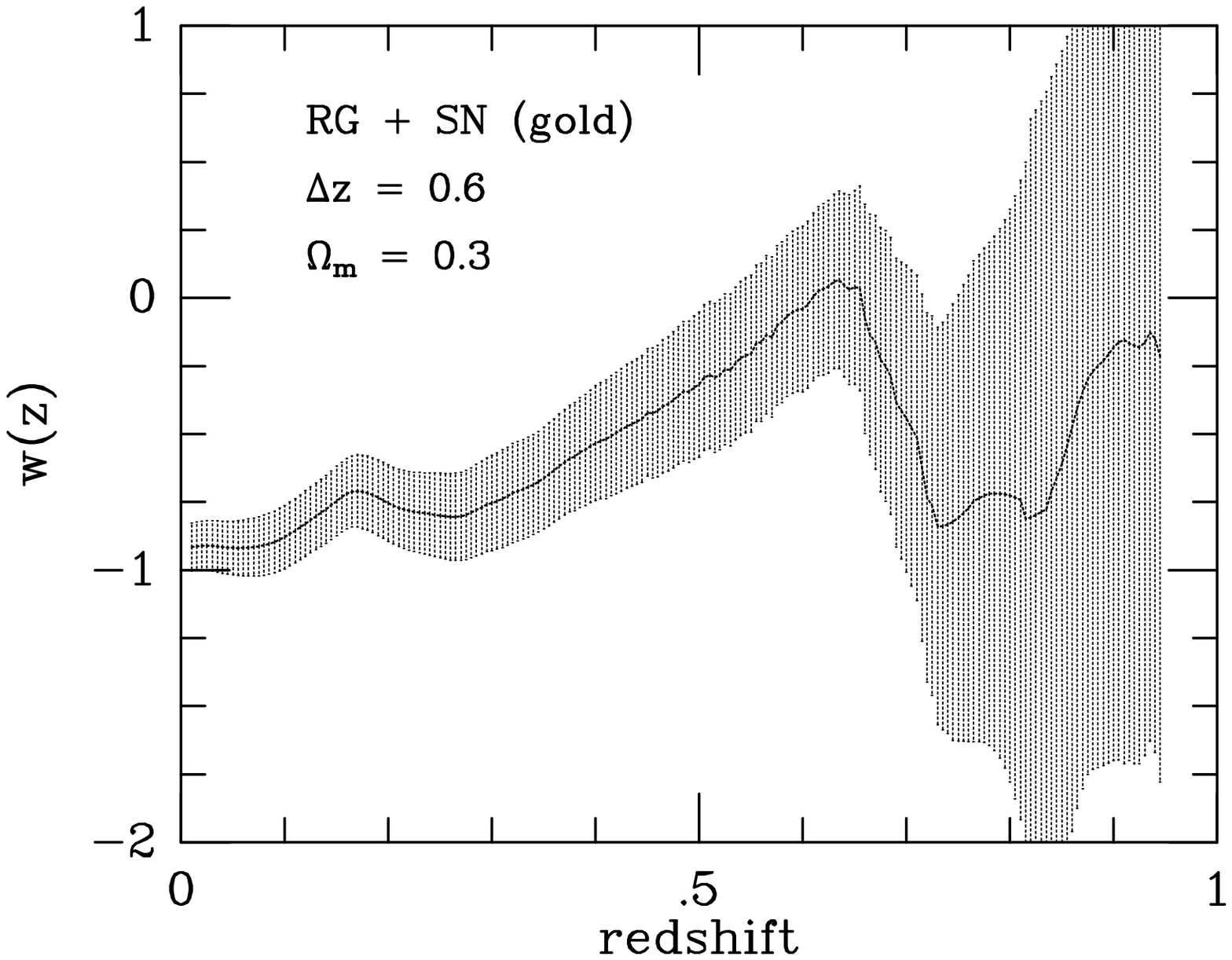}}
\caption{The derived values of the dark energy equation of state
parameter $w(z)$, obtained with window function of width
$\Delta z = 0.6$.
This derivation of $w(z)$ requires of theory of gravity
and the value of $\Omega_0$;
GR has been adopted here, and
$\Omega_0 = 0.3$ is assumed.  The present value
is $w_0 = -0.9 \pm 0.1$, consistent with
cosmological constant models.
}
\end{figure}

\section{Results and Conclusions}

The results presented here follow
those presented by Daly \& Djorgovski (2003, 2004), where more
details can be found.  The
data used here includes 20 radio galaxies (RG)
compiled by Guerra, Daly, \& Wan (2000)
and
the ``gold'' supernova (SN) sample
compiled by Riess et al. (2004).  We note that in the redshift
interval where the two sets of coordinate distances (RG and SN)
overlap, the agreement is excellent, suggesting that neither one
is affected by some significant bias, and allowing us to combine
them for this study.

Measurements of luminosity distances and angular size distances
are easily converted to coordinate distances, $y(z)$.
Using some robust numerical differentiation method,
these can be used to determine the
first and second derivatives as functions of redshift, which
can be combined to determine
the dimensionless expansion and acceleration rates of the universe
as functions of redshift, and the pressure, energy density, and
equation of state of the dark energy as functions of redshift,
as described above.

We see that the universe transitions from acceleration to deceleration
at a redshift of about 0.4 (consistent with determinations by
Daly \& Djorgovski 2003, Riess et al. 2004, and Alam, Sahni,
\& Starobinsky 2004); and our determination 
only depends upon the assumption
that the universe is homogeneous, isotropic, and spatially flat.
Assuming GR, we solve
for the pressure, energy density, and equation of state of the dark
energy.  Each is generally consistent with remaining constant to
a redshift of about 0.5 and possibly beyond, but determining their
behavior at higher redshifts is severely limited by the available data.

As more and better data become available, this methodology can be used
to determine the evolution of the dark energy properties and the
observed kinematics of the universe with an increasing precision
and confidence.

\section*{Acknowledgments}
This work
was supported in part by the U. S. National Science Foundation
under grants AST-0206002, and Penn State University (RAD),
and by the Ajax Foundation (SGD).  Finally, we acknowledge the great
work and efforts of many observers who obtained the valuable data
used in this study.

\end{document}